\documentclass[11pt,a4paper]{article}
\usepackage[utf8]{inputenc}
\usepackage[T1]{fontenc}
\usepackage{lmodern}
\usepackage[margin=2.5cm]{geometry}
\usepackage{setspace}
\usepackage{parskip}
\usepackage{graphicx}
\usepackage{booktabs}
\usepackage{enumitem}
\usepackage{hyperref}
\hypersetup{colorlinks=true,linkcolor=blue,citecolor=blue,urlcolor=blue}
\usepackage{fancyhdr}
\usepackage{caption}
\usepackage{amsmath,amssymb}

\usepackage[numbers]{natbib}

\begin{document}

\begin{titlepage}
  \thispagestyle{empty}
  \centering
  % \vspace*{1cm}

  {\Huge\bfseries Expanding Horizons \\[6pt]
   \large Transforming Astronomy in the 2040s\\
   \small White Paper in response to ESO\\ \par}

  \vspace{1cm}

  {\Large \textbf{How Mass Flows Through Accretion Discs:\\A Spectral–Timing Vision for the 2040s}\\\par}
  \vspace{1cm}

  \begin{tabular}{p{4.5cm}p{10cm}}
    \textbf{Scientific Categories:} &
      Time-domain, Stars, Binaries, Accretion \\[0.8em]

    \textbf{Submitting Author:} & Name: Simone Scaringi\\
    & Affiliation: Durham University (UK) \\
    & Email: \href{mailto:simone.scaringi@durham.ac.uk}{simone.scaringi@durham.ac.uk}
\\[0.8em]

    \textbf{Contributing authors:} &
    {\small
    Domitilla de Martino (INAF - OACN, IT), Anna F. Pala (ESO, Garching, DE), Andrea Sanna (Università degli Studi di Cagliari, IT), Paul Groot (Radboud University, NL; University of Cape Town \& SAAO, ZA), Kieran O'Brien (Durham University, UK),  Alessandro Ederoclite (CEFCA, ES), Noel Castro Segura (University of Warwick, UK), Deanne L. Coppejans (University of Warwick, UK), Krystian I\l{}kiewicz (CAMK PAN, PL), Piergiorgio Casella (INAF - OAR, IT), David Buckley (South African Astronomical Observatory; University of Cape Town, ZA), Thomas Kupfer (University of Hamburg, DE), Nanda Rea (ICE-CSIC, ES), Meryem Kubra DAG (Durham University, UK), Yusuke Tampo (South African Astronomical Observatory; University of Cape Town, ZA), Siqi Zhang (Durham University, UK), Sian Ford (Northumbria University, UK), Martina Veresvarska (ICE-CSIC, ES), Graham Wynn (Northumbria University, UK)
    
    }
    \\

  \end{tabular}

\vspace{1cm}

% ABSTRACT TO ONLY APPEAR ON THE ARXIV VERSION

  \textbf{Abstract:}

  \vspace{0.5em}
  \begin{minipage}{0.99\textwidth}
    \small
Understanding how mass and angular momentum flow through accretion discs remains a fundamental unsolved problem in astrophysics. Accreting white dwarfs offer an ideal laboratory for addressing this question: their variability occurs on accessible timescales of seconds to minutes, and their optical spectra contain continuum and emission-line components that trace distinct disc regions. Broad-band timing studies have revealed time-lags similar to those observed in X-ray binaries and active galactic nuclei, suggesting propagating fluctuations and possible coupling to an inner hot flow. However, the blending of line and continuum light in broad filters prevents a physical interpretation of these signals. The 2040s will bring an unprecedented number of disc-accreting systems discovered by Rubin-LSST, space-based gravitational-wave observatories, and third-generation ground and space-based detectors. To extract disc physics from these sources, high-cadence optical spectral-timing, simultaneously resolving continuum and individual lines, is essential. Such measurements would directly map how variability propagates through discs, determine how the outer disc responds to changes in the inner flow, and test whether accretion physics is scale-invariant from white dwarfs to supermassive black holes. This white paper outlines the scientific motivation and observational capabilities required to realise this vision. It highlights the opportunity for ESO to enable a transformative new window on accretion physics in the coming decade.
\end{minipage}

\end{titlepage}

\section{Introduction and Background}
\label{sec:intro}

Accretion discs power a vast range of astrophysical systems, from newborn stars and interacting white dwarfs to X-ray binaries and active galactic nuclei. Despite this ubiquity, the fundamental physical mechanism responsible for transporting angular momentum through discs remains unresolved. Magnetorotational turbulence, spiral shocks, thermal–viscous instabilities, magnetic winds, and interactions with magnetospheres have all been proposed as dominant transport channels, yet none can be confirmed empirically across all systems. A unified understanding of disc accretion requires direct, time-resolved measurements of how mass and energy propagate through real discs operating under different physical conditions, influenced by magnetic fields and variations in mass transfer rate.

Accreting white dwarfs (AWDs) offer a uniquely powerful laboratory for addressing this problem. They are nearby, bright, and evolve on dynamical and thermal timescales that are directly accessible to optical observations. Their spectra contain both continuum emission and a rich series of hydrogen and helium lines, each arising from distinct regions of the disc or its boundary layer, and/or the hot inner flow. This makes AWDs the only class of accreting objects in which we can attempt to map disc structure and variability simultaneously through time-resolved spectroscopy at optical wavelengths.

Optical timing measurements have revealed Fourier-dependent time lags between broad photometric bands in AWDs, reminiscent of the spectral–timing behaviour seen in X-ray binaries and supermassive black holes [1]. These lags are promising indicators of propagating accretion-rate fluctuations, yet they are currently difficult to interpret because the broad filters mix line and continuum contributions. To determine where these lags originate, whether in the continuum-emitting disc, in specific line-forming regions, or in an inner hot flow, and how these lags are affected by changes in the mass transfer rate occurring on longer timescales (months to years), we require time-resolved spectroscopy with both high cadence and high signal-to-noise on a large sample of systems.

By the late 2030s, this need will become urgent. Next-generation time-domain surveys will deliver an unprecedented volume of variable and transient sources. Rubin-LSST, for example, will identify millions of variable compact binaries annually. At the same time, space-based gravitational-wave interferometers such as \textit{LISA} will detect tens of thousands of interacting white dwarf binaries, many of which exhibit optical variability produced by disc processes. Third-generation gravitational-wave detectors, including the Einstein Telescope, will discover neutron star mergers whose electromagnetic emission evolves on minute-to-hour timescales. Yet no existing facility, nor any now under construction, is capable of providing the rapid, repeated, high-cadence spectroscopy needed to capture these systems in the time domain and disentangle the physics driving their accretion flows.

This white paper outlines the key scientific questions that will shape accretion disc research in the 2040s and describes the observational and technological capabilities required to address them. Although the context arises from AWDs, the implications extend across all mass scales: solving the accretion problem in white dwarfs will directly test our understanding of accretion physics in X-ray binaries and AGN and possibly compact object accretion during merger events, as well as providing the only empirical bridge between optical time domain and high-energy spectral–timing studies.

%-------------------------------------------------------------

\section{Open Science Questions in the 2040s}
\label{sec:openquestions}

The coming decade will allow, for the first time, direct measurement of how fluctuations in mass accretion rate propagate through a disc. Understanding this process requires identifying where variability is generated, how it moves through the disc, and how different emitting regions respond to it. AWDs, owing to their accessible timescales and spectroscopic richness, provide a tractable environment in which this mapping can be carried out in detail.

A first major question concerns the mechanisms that transport angular momentum through discs. In the classical picture, local viscous stresses redistribute angular momentum outward, allowing matter to spiral inward. However, it remains unclear whether magneto-rotational turbulence operates efficiently in the largely neutral discs of quiescent dwarf novae; whether spiral shocks, tidal forcing, or disc warping play a more dominant role; or whether magnetically driven winds extract angular momentum vertically rather than radially. Each of these mechanisms imprints a distinct causal structure on the disc's variability. For example, in a model dominated by propagating fluctuations [2,3], slow variations generated at large radii should progressively lead faster variations originating in the inner disc, giving rise to characteristic time lags between spectral components associated with those regions. Measuring these lags as a function of Fourier frequency in both continuum and individual emission lines, therefore, provides a direct probe of the physical processes governing angular-momentum transport.

A second open question is how the disc couples to any inner hot flow. In X-ray binaries and AGN, the inner accretion flow is believed to transition between optically thick and optically thin states, producing reverberation signals and reflection spectra that have become central to high-energy spectral–timing [4,5]. Whether AWDs host analogous inner structures remains unknown, though intermittent mode switching [6], disc truncation [7], and magnetically gated accretion [8] in several systems suggest that they may. Only simultaneous continuum and line timing can reveal whether the disc responds to irradiation from a variable central source or whether the variability originates locally within the disc. If reverberation-like behaviour can be detected in optical emission lines, it would be possible to map the absolute scale, geometry, and state-dependence of the inner disc in ways not previously possible.

A third central theme for the 2040s is the universality of accretion physics across all mass scales. There is growing evidence that accreting systems, from young stellar objects to supermassive black holes, exhibit common variability correlations, including a mass-scaling of characteristic break frequencies in their power spectra [9]. AWDs inhabit the missing size decade between X-ray binaries and AGN, yet their accretion flows operate at optical rather than X-ray wavelengths. If their spectral–timing behaviour mirrors that seen at higher energies, then the processes that set disc structure and variability are likely universal. This would represent a significant conceptual unification of accretion physics across many orders of magnitude in mass and size. Furthermore, the large populations of compact binaries that \textit{LISA} will reveal offer the prospect of connecting accretion-flow structure to binary parameters measured independently from gravitational waves.

Finally, a range of extreme or rare accretion phenomena, such as magnetically gated bursts [8], thermonuclear micronovae [10], quasi-periodic oscillations [11], and rapid state transitions [6], remain almost entirely unexplored in the spectral–timing domain. Many of these events evolve on minute-to-hour timescales and often occur unpredictably. To capture them, we require facilities capable of both long-term monitoring and rapid reactive spectroscopy, with the ability to concentrate or distribute observing power depending on the rarity or importance of the event. The discovery space for such phenomena will grow dramatically in the next decade, and their detailed spectroscopic timing will be necessary to understand the physics of these events.

%-------------------------------------------------------------

\section{Technology and Data Handling Requirements}
\label{sec:tech}

Answering these questions requires a transformation in how optical spectroscopy is acquired and processed. Traditional long-exposure spectroscopy is optimised for static or slowly varying sources. In contrast, the physics of accretion discs is encoded in rapid, stochastic variability that demands continuous, high-cadence spectral information. The core capability needed for the 2040s is therefore the routine acquisition of uninterrupted, read-out noise-free, photon-resolved optical spectra in the UV-opt-nIR (0.3-2.2$\mu$m) range, with a resolution of $R\sim$5000-10000 and sufficiently high temporal resolution to resolve disc variability on timescales of seconds.

Next-generation photon-counting detectors provide a pathway to this regime. By time-tagging individual photons and recording their spectral information without read-out noise, these detectors allow exposures to be defined entirely in software. The same dataset can be analysed at sub-second cadence to probe propagating fluctuations, or at minute cadence to detect slow structural evolution in the disc. This flexibility is essential for mapping lag–frequency spectra and determining the causal relationship between continuum and line emission. In addition, high duty cycle and continuous acquisition eliminate the dead time that currently obscures variability on the shortest accessible timescales.

Equally important is the ability to observe a sample of systems, as our hypotheses on accretion physics must be tested across a population of objects spanning a range of mass-transfer rates, disc geometries, and evolutionary states. A facility capable of obtaining high-cadence spectra for large numbers of AWDs regularly would enable unprecedented statistical studies, reveal rare behaviours, and capture systems in multiple accretion states. For faint systems, the capacity to increase effective collecting area, whether through modular telescope architectures or coordinated ensembles, will be critical, particularly for gravitational-wave sources whose optical counterparts may be intrinsically faint.

Real-time data analysis will also be essential. The volume of photon event data generated by continuous spectral–timing requires automated pipelines capable of producing power spectra, coherence measurements, and lag spectra with minimal latency. Such pipelines must be integrated with adaptive scheduling systems that can respond dynamically to unusual variability or external triggers. In the context of a crowded alert environment dominated by large area synoptic photometric surveys and gravitational-wave facilities, cognitive scheduling and machine-learning-based prioritisation will be necessary to deploy observing resources efficiently.

Finally, robust data formats and archival systems are required to preserve the time-domain integrity of these datasets. 
Event-based data structures, coupled with scalable storage and fast-access indexing, will allow the broader community to reanalyse spectral–timing information using future methods yet to be developed. Cross-referencing with photometric surveys, catalogue data, and gravitational-wave detections will enable entirely new forms of joint analysis.

% \bigskip
%In summary, addressing the fundamental question of how mass flows through accretion discs requires a new generation of optical time-domain spectroscopic capabilities. By enabling continuous, photon-counting spectroscopy across large samples of interacting binaries, ESO has the opportunity to open a discovery space that will unify accretion physics from white dwarfs to black holes, and anchor electromagnetic and gravitational-wave views of compact object accretion in the 2040s and beyond.

\section*{References}
\footnotesize
\textbf{[1]} Scaringi, S. et al. 2013, MNRAS, 431, 2535–2541, doi:10.1093/mnras/stt347; 
\textbf{[2]} Ingram, A. \& Done, C. 2011, MNRAS, 415, 2323–2335, doi:10.1111/j.1365-2966.2011.18860.x; 
\textbf{[3]} Scaringi, S. 2014, MNRAS, 438, 1233–1241, doi:10.1093/mnras/stt2270; ~
\textbf{[4]}~Arévalo, P.~\& Uttley, P. 2006, MNRAS, 367, 801–814, doi:10.1111/j.1365-2966.2006.09989.x; 
\textbf{[5]} Done, C., Gierliński, M. \& Kubota, A. 2007, A\&A Rev., 15, 1–66, doi:10.1007/s00159-007-0006-1; 
\textbf{[6]} Scaringi, S. et al. 2022, Nature Astronomy, 6, 98–102, doi:10.1038/s41550-021-01494-x; 
\textbf{[7]} Dubus, G. \& Babiusiak, C. 2024, A\&A, 683, A247, doi:10.1051/0004-6361/202348510; 
\textbf{[8]} Scaringi, S. et al. 2017, Nature, 552, 210–213, doi:10.1038/nature24653; 
\textbf{[9]} Scaringi, S. et al. 2015, Science Advances, 1, e1500686, doi:10.1126/sciadv.1500686; 
\textbf{[10]} Scaringi, S. et al. 2022, Nature, 604, 447–450, doi:10.1038/s41586-022-04495-6; 
\textbf{[11]} Veresvarska, M. et al. 2024, MNRAS, 534, 3087–3103, doi:10.1093/mnras/stae2279.

% \begingroup
%   \setlength{\parskip}{0pt}
%   \setlength{\itemsep}{0pt}
%   \footnotesize
%   \bibliographystyle{unsrtnat}
%   \bibliography{bibliography}
% \endgroup

\end{document}